\documentclass[prd,aps,12pt,showpacs,showkeys,nofootinbib]{revtex4-1}
\begin{document}

\title{Cosmic inflation and big bang interpreted as explosions}

\author{E.~Rebhan\footnote{rebhan@thphy.uni-duesseldorf.de}}
\affiliation{Institut f{\"u}r Theoretische Physik,\\
   Heinrich--Heine--Universit{\"a}t,\\
   D--40225 D{\"u}sseldorf, Germany}

\begin{abstract}
  It has become common understanding that the recession of galaxies
  and the corresponding redshift of light received from them can only
  be explained by an expansion of the space between them and us. In
  this paper, for the presently favored case of a universe without
  spatial curvature, it is shown that this interpretation is
  restricted to comoving coordinates. It is proven by construction
  that within the framework of general relativity other coordinates
  exist in relation to which these phenomena can be explained by a
  motion of the cosmic substrate across space, caused by an
  explosion like big bang or by inflation preceding an almost big
  bang. At the place of an observer, this motion occurs without any
  spatial expansion. It is shown that in these
  ``explosion coordinates'' the usual redshift comes about by a
  Doppler shift and a subsequent gravitational shift. Making use of
  this interpretation, it can easily be understood why in comoving
  coordinates light rays of short spatial extensions expand and thus
  constitute an exemption from the rule that small objects up to the
  size of the solar system or even galaxies do not participate in the
  expansion of the universe. It is also discussed how the two
  interpretations can be reconciled with each other.
\end{abstract}


\maketitle

\section*{Introduction}

After its discovery, the redshift of light from far away galaxies was
first explained in the context of special relativity and attributed to
a Doppler shift caused by an outward flight of the galaxies in a
preexisting invariable space. It was first observed by V.~M.~Slipher
in the years since 1912 (see e.g.~\cite{Slipher}), and in 1918
C.~Wirtz interpreted it as being due to a general recessive motion of
galaxies~\cite{Wirtz}. Use of the Doppler formula led to good
agreement with observational results and revealed the validity of the
Hubble law that, in fact, was first derived in 1927 from general
relativity by G.~Lemaitre~\cite{Lemaitre}. Only in 1929 it was
formulated by E.~Hubble in the context of astronomical
observations~\cite{Hubble}.  Hubble's interpretation of the
cosmological redshift as a Doppler shift became the generally accepted
view, and in accordance with it the big bang was considered as a giant
explosion.

The successful general relativistic formulation of the basic
cosmological equations in comoving coordinates, first by
A.~Friedman~\cite{Friedman} in 1922, later independently by
Lemaitre~\cite{Lemaitre} in 1927, and 1936 supplemented by a rigorous
derivation of the corresponding metric by
H.~P.~Robertson~\cite{Robertson} and A.~G.~Walker~\cite{Walker}, brought
a second interpretation into play. According to this the recession of
galaxies from a distant observer is not caused by a motion relative to
their spatial environment but rather by an expansion of the space
between them and the observer. Correspondingly inflation and the big
bang are sometimes denoted as space explosions.

For a long time the two interpretations coexisted because in those
days the most distant observable galaxies gave rise to very small
redshifts $z$ only, and at small $z$ the velocity - redshift relation
is the same for both interpretations. However, at least since 1998 in
the context of the Supernova Cosmology Project~\cite{Perlmutter} more
distant cosmic objects (supernovae) were observed and discrepancies
between the two interpretations became evident, the emphasis has
completely shifted to spatial expansion. By now even for small $z$ the
recession of galaxies is exclusively attributed to spatial expansion
whereas the explosion perspective is completely ruled out. This has
become the standard doctrine in research articles (see
e.g.~\cite{Davis} and~\cite{Davis2}), in textbooks (see
e.g.~\cite{Farmer}--\cite{Peacock} and~\cite{Mukhanov}), and
in Wikipedia~\cite{wikipedia}, the interpretation as an explosion
sometimes being denoted as a \textit{popular} or \textit{common
  misconception} (see e.g.~\cite{Davis},
\cite{Peacock}--\cite{science} or~\cite{Mukhanov}, page~28). In the
world wide web a multitude of contributions arguing against the
interpretation as an explosion can be found (see
e.g.~\cite{www1}--\cite{www7}).

In 1934 W.~H.~McCrea and E.~A.~Milne~\cite{Crea-Milne} showed that the
Friedmann-Lemaitre equations describing the dynamics of the cosmic
scale factor $a(t)$ in general relativity follow in exactly the same
form from Newton's laws of motion and of gravity (more precisely: from
the corresponding fluid equations by Euler). In this so-called
Newtonian cosmology the galaxies are located at the positions
$\vec{r}(t) \,{=}\, [a(t)/a(t_0)]\,\vec{r}(t_0)$ where $\vec{r}$ is the
radius vector in a Euclidean space and $\vec{r}\,{=}\,0$ is our position in
the universe, and they move with the velocities $\vec{v} \,{=}\,
H(t)\,\vec{r}$ (where $H(t)\,{=}\,\dot{a}(t)/a(t)$) relative to a
preexisting and invariable space. Near the origin the velocities of
galaxies and the gravitational field produced by them are so small
that the Newtonian equations of motion should asymptotically coincide
with equations obtainable from general relativity. This suggests that
also in general relativity there should exist coordinates in which the
galaxies are moving across space. 

In this paper, for the presently favored case of a universe without
spatial curvature, it will be shown that the seemingly contradictory
interpretations, motion relative to an invariable space and recession
due to expansion of space, are both possible and not in contradiction
with each other. It will become apparent that they are relative views
restricted to special sets of coordinate systems which are related to
each other by simple transformations.

It is clear that in the transition from one coordinate system to
another also the interpretation of physical phenomena can
change. Since in general relativity a huge variety of different
coordinate systems is available, it may at first glance appear almost
trivial to find a coordinate system which allows for a prescribed
interpretation -- motion across space instead of expansion of space in
our case. There are, however, several restrictions making things less
trivial. In order to remain within the framework of general
relativity, the signature of the metric of all admissible coordinate
systems must be the same.%
\footnote{In this paper SI units are used whence the speed of light
  is~$c$. Furthermore the signature of the metric is (+, --, --,
  --).}%
~In addition, conditions for excluding radial expansion must be
imposed. This will lead us to a set of differential equations to be
solved under the observance of specific boundary conditions. It is not
obvious that solutions of this problems exist, and even when they
exist, it is still not granted that the corresponding metric has the
appropriate signature. A simple example may demonstrate the problem.

In Newtonian mechanics one can go from the Cartesian coordinates of an
inertial system to any other coordinates and still remain within the
framework of the theory, provided that pseudoforces are admitted.
Specifically one can go to a rigidly rotating coordinate system. The
fact that beyond a certain radius this system rotates with
superluminal velocity poses no problem in Newtonian mechanics. In
general relativity, however, rigidly rotating coordinates are not
admissible because in the superluminal region the metric assumes the
wrong signature. A second example is presented in
footnote~\ref{footnote}.

The main body of this paper consists of proving the existence of
explosion coordinates. This is achieved in a constructive way deriving
them by means of a transformation from Robertson-Walker
coordinates. In section~\ref{sec:Syst-wo} the determining differential
equations and boundary conditions are specified, formal solutions are
derived and properties of them as well as conclusions following from
them are discussed. In Section~\ref{sec:spec-case} explicit solutions
for the most important special cases of cosmic evolution are deduced,
thus bringing to an end the envisaged proof of existence. As already
mentioned the interpretation of physical phenomena like the redshift
of light from far away galaxies must be adjusted. It will be shown
that the latter can exactly be described by a Doppler shift and a
subsequent gravitational shift. In this context it will be discussed
why and how a spontaneous Doppler shift at the place of emission can
be reconciled with the fact that in comoving coordinates ``...~the
increase of wavelength from emission to absorption of light does not
depend on the rate of change of [the cosmic scale factor] $a(t)$ at
the times of emission or absorption, but on the increase of $a(t)$ in
the whole period from emission to absorption''~\cite{Weinberg}.  In
section~\ref{sec:spec-case} it is also discussed to what extent the
cosmic explosion provided by inflation or a big bang can be compared
with usual explosions. Section~\ref{sec:applic} deals with some
applications and extensions.

It is clear that the purpose of this paper cannot be the replacement
of the usual approach with Robertson Walker coordinates and the
corresponding interpretation (also, see Section~\ref{sec:conclusion}),
especially since it starts off with solutions obtained in them. Rather
the paper is meant to provide a supplementation which might even turn
out to be useful in specific cases.

\section{Systems without radial expansion -- general theory}
\label{sec:Syst-wo}

In the case of a spatially uncurved universe the square of the line
element in (comoving) Robertson-Walker coordinates is
\begin{equation}                                                \label{eq:1}
  ds^{2} = c^{2} dt^{2} - a^{2}(t) \,\left(dr^{2} + r^{2} d\Omega \right)
 \quad\quad \mbox{with}\quad 
  d\Omega = d\vartheta^{2} + \sin^{2} \!\vartheta \,d\varphi^{2}\,.
\end{equation}
($r$ is dimensionless and $a(t)$ has the dimension of a length.) The
radial expansion of the universe is expressed by the time dependence
of length elements, e.g.  $dl_r\,{=}\,a(t)\,dr$ in radial direction. The
underlying coordinates $t,r,\vartheta$ and $\varphi$ are called
\textit{expansion coordinates} in this paper.  We are looking for a
transformation $t,r,\vartheta,\varphi\to\tau,\rho,\vartheta,\varphi$
to new coordinates $\tau,\rho, \vartheta$ and $\varphi$, the
\textit{explosion coordinates}, for which the square of the line
element is given by
\begin{equation}                                                \label{eq:2}
  ds^{2} = c^{2} g_{00}(\rho,\tau)\, d\tau^{2} - d\rho^{2} + g_\Omega(\rho,\tau)
  \,d\Omega\,.
\end{equation}
(In contrast to $r$, $\rho$ has the dimension of length.) The radial
length element $dl_\rho\,{=}\,d\rho$ is time independent whence in explosion
coordinates there is no radial expansion. As we shall see instead of
this there is a radial motion. Since $\vartheta$ and $\varphi$ remain
unchanged, we have
\begin{equation}                                                \label{eq:3}
  t = t(\rho,\tau)\,, \quad\quad r = r(\rho,\tau)\,. 
\end{equation}
Other than in Newtonian cosmology the space is not completely
invariable but may involve angular expansion because $g_\Omega$ can be
time dependent. A corresponding expansion of volumes will be discussed
at the end of this section.

\subsubsection{Derivation of explosion coordinates
  and  corresponding metric}
From eqs.~(\ref{eq:3}) we get
\begin{displaymath}
  dt = t_\rho d\rho +t_\tau d\tau \,, \quad\quad 
  dr = r_\rho d\rho +r_\tau d\tau
\end{displaymath}
where $ t_\rho $ denotes the partial derivative of the function
$t(\rho,\tau)$ with respect to $\rho$ etc.. With this, from
eq.~(\ref{eq:1}) we obtain
\begin{displaymath}
  ds^{2} = \big(c^2t^2_\rho-a^2 r^2_\rho\big)\, d\rho^2
         + \big(c^2t^2_\tau -a^2 r^2_\tau \big)\, d\tau^2
         +2\big(c^2 t_\rho t_\tau -a^2r_\rho r_\tau \big)\, d\tau^2
         +\dots\,.
\end{displaymath}
In order for this to assume the form of eq.~(\ref{eq:2}), the
equations
\begin{equation}                                          \label{eq:4}
  c^2 t_\rho t_\tau = a^2r_\rho r_\tau \,, \quad\quad
  c^2t^2_\rho - a^2 r^2_\rho = -1 \,, \quad\quad
  g_{00}=t^2_\tau-\frac{a^2 r^2_\tau}{c^2} \,, \quad\quad
  g_\Omega = - a^2\,r^2
\end{equation}
must be fulfilled, or, equivalently
\begin{equation}                                         \label{eq:5}
  r_\rho = \pm\frac{\sqrt{1+c^2t^2_\rho}}{a(t)}\,, \quad\quad
   r_\tau =  \pm\frac{c^2 t_\rho t_\tau}{a(t)\,\sqrt{1+c^2t^2_\rho}}\,.
\end{equation}
In order to make the results in expansion  and explosion coordinates
comparable we impose the following conditions: 1. the origins of the
two coordinate systems permanently coincide, and 2. the coordinate
times at the origins are the same,
\begin{equation}                                               \label{eq:9} 
  1.\quad 
   \rho = 0 \quad\mbox{for}\quad\quad r = 0 \,,\quad\quad\quad\quad
  2.\quad
  t=\tau  \quad \mbox{at} \quad\quad \rho=0\,.
\end{equation}
 
Eqs.~(\ref{eq:5}) are two equations only for the determination of the
four derivatives $ r_\rho, r_\tau, t_\rho$ and $t_\tau$, since as soon
as the transformation functions (\ref{eq:3}) have been determined from
eqs.~(\ref{eq:5}), the last two of the eqs.~(\ref{eq:4}) only serve
for the evaluation of $g_{00}$ and $g_\Omega$. For eqs.~(\ref{eq:5})
to have solutions, the integrability condition
$r_{\rho\tau}\,{=}\,r_{\tau\rho}$ must be satisfied. The evaluation of it
results in a nonlinear second order differential equation for the
function $t(\rho,\tau)$,
\begin{equation}                                          \label{eq:6}
  a(t)\,t_{\rho\rho}+\frac{\dot{a}(t)}{c^2}\,\big(c^2t^2_{\rho}+1\big)=0\,.
\end{equation}
Once a solution $t(\rho,\tau)$ is found, $r(\rho,\tau)$ can be
determined from eqs.~(\ref{eq:5}).  After multiplication with
$2a\,t_\rho$ eq.~(\ref{eq:6}) becomes
\begin{displaymath}
  2a^2(t)\, t_\rho t_{\rho\rho} +2a(t)\,\dot{a}(t)\,t^3_{\rho}
  +\frac{2a(t)\,\dot{a}(t)}{c^2}\,t_\rho=\frac{\partial}{\partial\rho}
  \left[a^2(t)\big(t^2_\rho+1/c^2\big)\right]=0\,.
\end{displaymath}
This equation can be integrated once to yield
\begin{equation}                                            \label{eq:7}
  t_\rho =\pm\frac{1}{c}\,\sqrt{F(\tau)/a^2(t)-1}
\end{equation}
and once more to yield%
\vskip -\baselineskip
\begin{equation}                                               \label{eq:8} 
 \rho = G(\tau)\pm \left. c\int_0^t\frac{dt'}{\sqrt{F(\tau)/a^2(t')-1}}
 \right|_{\tau=const}
\end{equation}
where $F(\tau)$ and $G(\tau)$ are integration functions. (Note that
the algebraic sign in eqs.~(\ref{eq:7}) and (\ref{eq:8}) can be chosen
independently from that in the eqs.~(\ref{eq:5}).) The latter can be
chosen in such a way that additional conditions are satisfied.  Using
the second of the conditions~(\ref{eq:9}), from eq.~(\ref{eq:8}) we get
\begin{equation}                                               \label{eq:8a} 
 G(\tau) = \mp \left. c\int_0^\tau\!\frac{dt'}{\sqrt{F(\tau)/a^2(t')-1}}
 \right|_{\tau=const}\,.
\end{equation}
Inserting eq.~(\ref{eq:7}) and the second of the eqs.~(\ref{eq:5}) in
the third of the eqs.~(\ref{eq:4}) yields
\begin{equation}                                              \label{eq:8b} 
  g_{00} = \frac{ t^2_\tau }{1+c^2\,t^2_\rho} = \frac{a^2(t)\,t^2_\tau}{F(\tau)}\,.
\end{equation}
Since under observance of both conditions (\ref{eq:9}) the origins of
the two coordinate systems permanently coincide and since
$t\,{=}\,\tau$ there, also the metric times must be the same
there. (Although this statement needs no separate proof, it will later
be validated as a test in special cases.) As a result of this from
eqs.~(\ref{eq:1}) and (\ref{eq:2}) we get
\begin{equation}                                           \label{eq:8bb}
  g_{00}(\rho=0,\tau)\equiv 1\,.
\end{equation}
With this and $t_\tau\,{=}\,1$ for $\rho\,{=}\,0$ since $t\,{=}\,\tau$ for $\rho\,{=}\,0$,
from eq. (\ref{eq:8b}) we finally obtain
\begin{equation}                                              \label{eq:8c} 
   F(\tau) = a^2(\tau)\,.
\end{equation}
Here, $a(\tau)$ stands for the function $a(t)$ with $t$ replaced by
$\tau$. With this and eq.~(\ref{eq:8a}) eq.~(\ref{eq:8})
becomes%
\vskip -1.5\baselineskip
\begin{equation}                                               \label{eq:8d} 
 \rho = \pm \left. c\int_\tau^t\frac{dt'}{\sqrt{a^2(\tau)/a^2(t')-1}}
 \right|_{\tau=const}\,.
\end{equation}
This equation implicitly determines the function $t \,{=}\, t(\rho,\tau)$.
Inserting eqs.~(\ref{eq:8c}) and (\ref{eq:7}) in the first of the
eqs.~(\ref{eq:5}) yields
\begin{equation}                                         \label{eq:rrho} 
  r_\rho = \pm\frac{a(\tau)}{a^2(t)} \quad\quad\mbox{and}\quad\quad
  r = R(\tau) \pm \,a(\tau)\left.\int_0^\rho\frac{d\rho'}
    {a^2(t(\rho',\tau))}\right|_{\tau=const}\,,
\end{equation}
where $R(\tau)$ is an arbitrary integration function. From the first
of the conditions (\ref{eq:9}) it follows that $R(\tau)\,{=}\,0$, so finally
we get
\begin{equation}                                           \label{eq:8e} 
   r(\rho,\tau) = \left.a(\tau)\int_0^\rho\frac{d\rho'}{a^2(t(\rho',\tau))}
   \right|_{\tau=const}\,.
\end{equation}
(The minus sign could be excluded since we must have $r>0$.)

The eqs.~(\ref{eq:8d}) and (\ref{eq:8e}) determine the solutions of
the eqs.~(\ref{eq:4}) for the transformation functions $r(\rho,\tau)$
and $t(\rho,\tau)$ only implicitly and require the knowledge of a
solution $a(t)$ of the Friedman-Lemaitre equations. Since according to
eq.~(\ref{eq:8d}) their range of validity is restricted to
$a(\tau)\geq a(t)$, it is at this point not yet clear whether for each
$a(t)$ also explicit solutions exist and in addition lead to the right
signature of the metric. Therefore, the proof of existence of
solutions is not yet complete. Their determination will in general
require numerical methods. However, in special cases also analytical
solutions can be obtained. In section~\ref{sec:spec-case} the full
solution for two relatively simple but nevertheless representative
cases will be derived.

\subsubsection{Flight velocity of galaxies}

In explosion coordinates the radial position of a galaxy (or an element
of the cosmic substrate) is given by $r(\rho,\tau)\,{=}\,const$, whence
$\rho\,{=}\,\rho(\tau)$ and
\begin{equation}                                          \label{eq:8f} 
  \dot{\rho}(\tau) = -\frac{r_\tau}{r_\rho} = 
  \pm\frac{c\,a^2(t)\,t_\tau\,\sqrt{a^2(\tau)/a^2(t)-1}}{a^2(\tau)}\,.
\end{equation}
(For the last step eqs.~(\ref{eq:5}), (\ref{eq:7}) and (\ref{eq:8c})
were used.) This means that the galaxies are moving across
space. $\dot{\rho}(\tau)$ is the coordinate velocity, and from it the
physical velocity $v$ is obtained according to%
\vskip -1.5\baselineskip
\begin{equation}                                         \label{eq:8ff} 
  v = \frac{\sqrt{g_{\rho\rho}}\;d\rho}{\sqrt{g_{00}}\;d\tau} 
  =\frac{d\rho}{\sqrt{g_{00}}\,d\tau} = 
    \frac{\dot{\rho}(\tau)\;a(\tau)}{a(t)\,t_\tau}\,,
\end{equation}
where at last eqs.~(\ref{eq:8b}) and (\ref{eq:8c}) were
used. Insertion of eq.~(\ref{eq:8f}) finally yields
\begin{equation}                                            \label{eq:8g}
  v = c\,\sqrt{1-a^2(t)/a^2(\tau)}\,.
\end{equation}
Thereby a freely disposable sign was chosen such that $v$ is positive.

\subsubsection{Redshift of light from far away galaxies}
1. In expansion coordinates, the redshift $z$ of light emitted from a
galaxy at the radius $r_{em}$ at the time $t_{em}$ and received at the
origin $r\,{=}\,0$ at the time $t_0$ is given by
\begin{equation}                                            \label{eq:8h}
  z+1 = \frac{\lambda_0}{\lambda_{em}} = \frac{a(t_0)}{a(t_{em})} 
\end{equation}
where $\lambda_{em}$ or $\lambda_0$ is the wavelength of the light at
emission or reception respectively. The radial propagation of light is
described by $c\,dt\,{=}\,a(t)\,dr$  whence
\begin{equation}                                          \label{eq:8hh}
  r_{em}= c\int_{t_{em}}^{t_0}\,\frac{dt'}{a(t')}\,.
\end{equation}
With this equation the time $t_0$ of reception can be expressed in
terms of the time $t_{em}$ and the location $r_{em}$ of emission.

2. In explosion coordinates for an observer fixed at the distance
$\rho\,{=}\,\rho_{em}$ from the origin the light emitted from a galaxy
flying past him at the radial velocity~$v$ undergoes the longitudinal
Doppler shift
\begin{equation}                                      \label{eq:17-}
  \frac{\lambda_\rho}{\lambda_{em}}=\left(\frac{1+v/c}{1-v/c}\right)^{1/2}
  =\frac{1+v/c}{\sqrt{1-v^2/c^2}}\,.
\end{equation}
The physical velocity of eq.~(\ref{eq:8g}) is the one that must be
inserted in the Doppler formula, and because unlike the recessional
velocity $\dot{a}(t)\,r$ in expansion coordinates it cannot exceed
the speed of light, it is always in the range of validity of this
formula. With it eq.~(\ref{eq:17-}) becomes%
\vskip -1.5\baselineskip
\begin{equation}                                       \label{eq:17}
  \frac{\lambda_\rho}{\lambda_{em}}=\frac{a(\tau)}{a(t)}\,\left(1+
    \sqrt{1-a^2(t)/a^2(\tau)}\right)\,,
\end{equation}
where $\tau\,{=}\,\tau_{em}$ or $t\,{=}\,t(\rho,\tau)\,{=}\,t_{em}$
are the times of emission in explosion  or expansion coordinates
respectively. On its way from $\rho\,{=}\,\rho_{em}$ to the observer
at $\rho\,{=}\,0$ the Doppler-shifted light experiences an additional
shift in the gravitational field acting in the system of
explosion coordinates. For simplicity we assume that this field is
time independent as is true for the special cases to be studied
below. Then, according to general relativity, the ratio of the
wavelength $\lambda_0$ observed at $\rho\,{=}\,0$ and $\tau=\tau_0$ to
the wavelength $\lambda_\rho$ of the Doppler-shifted light at
$\rho_{em}\to \rho$ and $\tau_{em}\to\tau$ is
\vskip -\baselineskip
\begin{equation}                                           \label{eq:8ccd} 
  \frac{\lambda_0}{\lambda_\rho} =
  \sqrt{\frac{g_{00}(\rho=0,\tau=\tau_0)}{g_{00}(\rho,\tau)}}\,.
\end{equation}
Making use of the eqs.~(\ref{eq:8b})--(\ref{eq:8c}), from this we
get
\begin{equation}                                       \label{eq:8cc}
   \frac{\lambda_0}{\lambda_\rho} = \frac{1}{\sqrt{g_{00}(\rho,\tau)}}
   =\frac{a(\tau)}{a(t)\,t_\tau}\,.
\end{equation}
Combining eqs.~(\ref{eq:17}) and (\ref{eq:8cc}) we finally
obtain
\begin{equation}                                       \label{eq:8dd}
   \frac{\lambda_0}{\lambda_{em}}=\frac{a^2(\tau)}{a^2(t)\,t_\tau}\,
   \left(1+\sqrt{1-a^2(t)/a^2(\tau)}\right)\,.
\end{equation}
Since observers at the origin use the same metric time in both
coordinate systems, they also measure the same frequency or redshift
of light that is emitted by far away galaxies. Therefore the result
(\ref{eq:8dd}) must be the same as the result provided by
eqs.~(\ref{eq:8h})--(\ref{eq:8hh}), and although not evident this
coincidence does not need proof. Nevertheless it is illuminating to
see it verified in the special cases considered in
section~\ref{sec:spec-case}.

\subsubsection{Volume expansion rate}
One could be tempted to assume that the elimination of radial
expansion in explosion coordinates leads to increased angular
expansion. Indeed the structure of $g_\Omega(\rho,\tau)$ in
explosion coordinates (see eq.~(\ref{eq:16}) for example) suggests
that there could still be a volume expansion due to a time dependence
of azimuthal distances. In order to clarify this issue we calculate an
expansion rate that involves both radial and angular properties, the
volume expansion rate $E=(d\Delta V/dT)/\Delta V$. In this $\Delta V$
is the volume of a spherical shell with infinitesimally small
thickness $\Delta\rho$, and $T=\sqrt{g_{00}}\,\tau$ is the time
measured on clocks in the system of explosion coordinates whence%
\vskip -0.5\baselineskip
\begin{equation}                                   \label{eq:defE}
  E=\frac{1}{\Delta V}\,\frac{d\Delta V}{\sqrt{g_{00}}\,d\tau}\,.
\end{equation}
According to eq.~(\ref{eq:2}) and the last of the eqs.~(\ref{eq:4})
\begin{displaymath}
  \Delta V = 4\pi\,g_\Omega\,\Delta\rho = 4\pi\,a^2(t)\,r^2\,\Delta\rho
\end{displaymath}
with $t\,{=}\,t(\rho,\tau)$ and $r \,{=}\, r(\rho,\tau)$. From this, for fixed
$\rho$ we obtain
\begin{displaymath}
  \frac{1}{\Delta V}\,\frac{d\Delta V}{d\tau} 
  =\frac{2\,[\dot{a}(t)\,r\,t_\tau+a(t)\,r_\tau]}{a(t)\,r}\,,
\end{displaymath}
and using eqs.~(\ref{eq:8b}) and (\ref{eq:8c}) we finally get
\begin{equation}                                       \label{eq:formE}
  E= \frac{2\,a(\tau)}{a(t)}\,
  \left(\frac{\dot{a}(t)}{a(t)}+\frac{r_\tau}{r\,t_\tau}\right)\,.
\end{equation}

For obtaining the expansion rate $E$ at $\rho\,{=}\,0$ we consider the volume
\begin{displaymath}
  V=4\pi\,\int_0^\rho a^2(t)\,r^2\,d\rho'
\end{displaymath}
of a sphere of radius $\rho$. Expanding around $\rho\,{=}\,0$ we have
\begin{displaymath}
  r(\rho,\tau) = r(0,\tau)+r_\rho(0,\tau)\,\rho + \dots
  =r_\rho(0,\tau)\,\rho + \dots
\end{displaymath}
because $ r(0,\tau)\,{=}\,0$ according to the first of the conditions
(\ref{eq:9}). Furthermore, according to the first of the
eqs.~(\ref{eq:rrho}) we have
\begin{displaymath}
  r_\rho(0,\tau) = \left.\frac{a(\tau)}{a^2(t)}\right|_{\rho=0} 
  = \frac{1}{a(\tau)}
\end{displaymath}
since $t\,{=}\,\tau$ for $\rho\,{=}\,0$ according to the second of the conditions
(\ref{eq:9}). In consequence
\begin{displaymath}
   r(\rho,\tau) = \frac{\rho}{a(\tau)} + \mathcal{O}(\rho^2)\,.
\end{displaymath}
Finally, we get
\begin{displaymath}
  a(t(\rho,\tau))=a(t(0,\tau))+ \mathcal{O}(\rho)
  =a(\tau)+ \mathcal{O}(\rho)\,.
\end{displaymath}
Altogether we have
\begin{displaymath}
  V = 4\pi\,\int_0^\rho \big[\rho'^2+\mathcal{O}(\rho'^3)\big]\,d\rho'
  =\frac{4\pi\,\rho^3}{3}+\mathcal{O}(\rho^4)\,.
\end{displaymath}
From this we get $dV/d\tau|_{\rho=0}\,{=}\,0$ and
$E|_{\rho=0}\,{=}\,0$, \textit{in explosion coordinates at the origin
  there is no volume expansion}.  Since in an expanding universe each
point can be chosen as the origin, in explosion coordinates at each
point for a local observer there is no expansion. The expansion around
points at some distance from the origin, described by equation
(\ref{eq:formE}), can therefore be regarded as virtual.

\section{Systems without radial expansion -- special cases}
\label{sec:spec-case}

For simplicity our analysis of special cases will be restricted to
situations in which the gravitational field is static in
explosion coordinates. Since Newtonian cosmology is a very good
approximation in the neighborhood of the origin $r\,{=}\,0$, the
corresponding conditions can be determined with it. According to the
Hubble law $v\,{=}\,H(t)\,r$ the accelerating (or decelerating) field,
equaling the gravitational field (an inflation field being included),
is
\begin{displaymath}
  \dot{v}(t) =  \dot{H}(t)\,r + H\,\dot{r}(t) = (\dot{H}(t)+H^2)\,r\,.
\end{displaymath}
It becomes time independent for%
\vskip -0.5\baselineskip%
\begin{displaymath}
  \dot{H}(t)+H^2 = \left\{
    \begin{array}{c}
      0\,\\ 
      \,A^2=const\\
      -A^2=const\,.      
    \end{array}
    \right.
\end{displaymath}
\vskip -0.4\baselineskip%
\noindent
In the first case, we obtain
\begin{displaymath}
  H(t) = \frac{\dot{a}(t)}{a(t)} = \frac{1}{t} 
  \quad\quad\rightarrow \quad\quad
  a(t) = \alpha \, t 
\end{displaymath}
where $\alpha\,{=}\,const$, i.e., the universe is expanding with constant
velocity.  In the second case two solutions $H(t)$ exist, firstly
$H\,{=}\,A$, i.e. a universe with constant expansion rate, and secondly
\begin{displaymath}
  H(t) = A\,\coth (A\,t) 
  \quad\quad \rightarrow \quad\quad 
  a(t) = a_0\sinh (A\,t) \,.
\end{displaymath}
In the third case we obtain 
\begin{displaymath}
  H(t) = A\,\tanh (A\,t) 
  \quad\quad \rightarrow \quad\quad 
  a(t) = a_0\cosh (A\,t) \,.
\end{displaymath}
In the following we shall only consider the cases of constant
expansion velocity and constant expansion rate.

\subsection{Universe with constant expansion velocity}
\label{sec:const-vel}

For a long time the recent evolution of the universe up to the present
state was considered to be best represented by a model that assumes
the domination of matter and exhibits a decelerated expansion with
$a(t)\sim t^{3/2}$. Only shortly before the end of the last century it
was detected that the expansion is rather slightly
accelerated~\cite{Perlmutter}. The case of a universe without
acceleration or deceleration, $a(t)\,{=}\,\alpha\,t$ with
constant~$\alpha$, lies in between, represents a fairly good
approximation -- which is best at the time of the transition from
decelerated to accelerated expansion -- and enables a full analytic
solution of our problem.

\subsubsection{Derivation of explosion coordinates}
With $a(t)\,{=}\,\alpha\,t$ the integral in eq.~(\ref{eq:8d})
can readily be evaluated and is given by
\begin{displaymath}
  \int_\tau^t\frac{dt'}{\sqrt{a^2(\tau)/a^2(t')-1}}=\frac{1}{2}\,
  \int_\tau^t\frac{dt'^2}{\sqrt{\tau^2-t'^2}}
  =-\sqrt{\tau^2-t^2}\,.
\end{displaymath}
With this, resolving eq.~(\ref{eq:8d}) with respect to t and choosing
a sign such that $t\geq 0$, we obtain%
\vskip -0.5\baselineskip
\begin{equation}                                       \label{eq:10}
  t=\frac{1}{c}\,\sqrt{c^2\tau^2-\rho^2}\,.
\end{equation}
With $a(t)\,{=}\,\alpha\,t$ and this result eq.~(\ref{eq:8e}) becomes
\begin{equation}                                           \label{eq:13}
  r= \alpha \tau \int_0^\rho \frac{d\rho'}{\alpha^2\,t'^2(\rho',\tau)}
  =\frac{\tau\,c^2}{\alpha}\,\int_0^\rho
  \frac{d\rho'}{c^2\,\tau^2-\rho'^2}
  =  \frac{c}{2\alpha}\,\ln\frac{c\tau+\rho}{c\tau-\rho}\,.
\end{equation}
The transformation
$t,r,\vartheta,\varphi\to\tau,\rho,\vartheta,\varphi$, providing a
line element of the form~(\ref{eq:2}), is given by eqs.~(\ref{eq:10})
and (\ref{eq:13}), and the inverse transformation is
\begin{equation}                                        \label{eq:15}
  \tau = t\,\cosh\frac{\alpha\,r}{c}\,,\quad\quad
  \rho = ct\,\sinh\frac{\alpha\,r}{c}\,.
\end{equation}
Inserting eqs.~(\ref{eq:10}) and (\ref{eq:13}) in the third and fourth
of the eqs.~(\ref{eq:4}) yields
\begin{equation}                                      \label{eq:16}
  g_{00} = 1\,,\quad\quad 
  g_\Omega = -\frac{c^2\tau^2-\rho^2}{4}\,
  \left(\ln\frac{c\tau+\rho}{c\tau-\rho}\right)^2\,.
\end{equation}
(In the derivation of the first equation, for the sake of brevity
eq.~(\ref{eq:8bb}) was used. One can easily make sure that the
same result is obtained without it.) Inserting the results
(\ref{eq:16}) in eq.~(\ref{eq:2}) yields%
\vskip -\baselineskip
\begin{equation}                                                \label{eq:16f}
  ds^{2} = c^{2} \, d\tau^{2} - d\rho^{2} -\frac{c^2\tau^2{-}\rho^2}{4}\,
  \left(\ln\frac{c\tau+\rho}{c\tau-\rho}\right)^2 \,d\Omega\,.
\end{equation}

\subsubsection{Flight velocity of galaxies and associated redshift}
The flight velocity of a galaxy at $r\,{=}\,const$ can be obtained from
eqs.~(\ref{eq:8g}) and (\ref{eq:10}) with
$a(t)\,{=}\,\alpha\,t$. Alternatively, from eqs.~(\ref{eq:15}) we
immediately get
\begin{equation}                                      \label{eq:x}
  \rho = c\,\tau\,\tanh\frac{\alpha\,r}{c}
  \quad\quad\mbox{and}\quad\quad
  v = \frac{\rho}{\tau} = c\,\tanh\frac{\alpha\,r}{c}\,.
\end{equation}

1. For expansion coordinates, with $a(t)\,{=}\,\alpha\,t$ the
eqs.~(\ref{eq:8h}) and (\ref{eq:8hh}) yield
\begin{displaymath}
  \frac{\lambda_0}{\lambda_{em}}  = \frac{t_0}{t_{em}}
  \quad\quad\mbox{and}\quad\quad
  r_{em}= \frac{c}{\alpha}
  \int_{t_{em}}^{t_0}\frac{dt}{t}=\frac{c}{\alpha}\ln\frac{t_0}{t_{em}}
\end{displaymath}
whence%
\vskip -1.2\baselineskip
\begin{equation}                                        \label{eq:16c}
  \frac{\lambda_0}{\lambda_{em}} = \mbox{e}^{\alpha r/c}\,.
\end{equation}

2. In explosion coordinates from eqs.~(\ref{eq:17-}) and (\ref{eq:x})
we get
\begin{equation}                                       \label{eq:17+}
  \frac{\lambda_\rho}{\lambda_{em}}=\left[\frac{1+\tanh(\alpha r/c)}
    {1-\tanh(\alpha r/c)}\right]^{1/2}= \;\mbox{e}^{\alpha r/c}\,.
\end{equation}
According to the first of the eqs.~(\ref{eq:16}) there is no
additional gravitational redshift of the light on its way from
$\rho\,{=}\,\rho_{em}$ to $\rho\,{=}\,0$,
\i.e. $\lambda_0/\lambda_\rho\,{=}\,1$, and therefore we obtain
\begin{equation}                                     \label{eq:18}
  \frac{\lambda_0}{\lambda_{em}}=\mbox{e}^{\alpha r/c}\,,
\end{equation}
exactly the same result as in expansion coordinates,
eq.~(\ref{eq:16c}). It should be noted, however, that, following from
eqs.~(\ref{eq:15}), (\ref{eq:1}) and (\ref{eq:16f}), in the two
coordinate systems the distance from the point of emission as well as
the time of emission differ from each other.

3. The case of constant expansion velocity is particularly suited for
demonstrating the compatibility of the two interpretations at issue,
because in explosion coordinates the redshift is completely due to the
Doppler effect. According to eqs.~(\ref{eq:x}) the spatial grid of
expansion coordinates $r$ is moving relative to that of
explosion coordinates $\rho$ at a speed that increases with the
distance from the origin. We now consider a light ray emitted at
$r\,{=}\,r_{em}$ and directed toward the origin.  From the viewpoint
of the system $S_{expl}$ of explosion coordinates, observers at rest
in the system $S_{expa}$ of expansion coordinates are moving relative
to the emitting galaxy at $r\,{=}\,r_{em}$ at a velocity that
increases with decreasing~$r$. Therefore in $S_{expl}$ they appear to
observe a Doppler shift that continuously increases as $r$ approaches
zero. This way an observer in $S_{expl}$ can readily understand why
for observers in $S_{expa}$ the redshift appears to be accumulated on
the way of light from emission to reception.

\subsubsection{View of an external observer}
According to the theory of chaotic inflation~\cite{Linde1,Linde2}, in
the case $k\,{=}\,0$ (no spatial curvature) considered in this paper a
Friedman-Lemaitre universe could be finite and embedded in an infinite
super-universe.  This so-called multiverse is filled with a ``foam''
of fluctuating quantum fields from which numerous universes of all
kinds can emerge by inflation or have already done so. It is
interesting to find out how the situation, so far considered from an
internal observer, would be seen by an external observer located at
some distance from our universe. 

Considerably simplifying the above model we consider a toy model of
the universe in which a finite spherical section of a
Friedman-Lemaitre universe is surrounded by a sufficiently large
bubble of true vacuum, uninfiltrated by any externally generated
gravitational fields. In contrast to the situation outside a
collapsing star, due to Birkhoff's theorem (time independence of all
metric coefficients in the vacuum surrounding a spherically symmetric
mass or energy distribution, see~\cite{Birkhoff} or e.g.~\cite{Wein})
the metric of the external space is not of the Schwarzschild type but
must be pseudo-Euclidean. The reason is that it has been of this kind
before the universe emerged by a creation out of nothing, a process
tolerated by general relativity when the big bang (or, rather, an
almost big bang) is preceded and triggered by an inflation field (dark
energy, inflaton).  Therefore, in polar spatial coordinates
$\rho,\vartheta$ and $\varphi$ the corresponding line element is
\begin{equation}                                             \label{eq:ex.1}
  ds^2 = c^2\,d\tau^2 - d\rho^2 - \rho^2\,d\Omega\,.
\end{equation}

The case $a(t)\,{=}\,\alpha\,t$ represents a fairly good approximation to
the state of our universe from the end of the period of matter
dominance until now, i.e. for the time interval \mbox{$t_0/2\leq t\leq
  t_0$} where $t_0$ is the present age of the universe. In this time
interval the pressure can be neglected and matter can be treated as
pressureless dust. The only forces acting on the matter elements are
gravitational forces (including the action of dark energy), and we can
therefore assume that in comoving coordinates the boundary of the
universe (which must lie well beyond our horizon) is at rest, $r \,{=}\, R \,{=}\,
const$, and is constituted by matter elements moving at the velocity
(\ref{eq:x}) with $r \,{=}\, R$.

Since according to eqs.~(\ref{eq:16f}) and (\ref{eq:ex.1}) the
external and internal coordinates are both not only Gaussian normal
but also employ the same radial metric,%
\footnote{\label{footnote}For a smooth connection with the external
  coordinates, as internal coordinates also standard coordinates
  ($ds^2 = g_{00}(\rho,\tau)\,c^2\,d\tau^2 +
  g_{\rho\rho}(\rho,\tau)\,d\rho^2 + \rho^2\,d\Omega$) could be
  envisaged. However, for \mbox{$a(t)\sim t$}, at some critical~$\rho$
  a coordinate singularity occurs, and for all $\rho>\rho_{crit}$
  standard coordinates do not exist. This provides another example
  that compatibility with the requirements of general relativity may
  prevent the existence of coordinates with specified properties.}%
~it is rather obvious that the velocity of the boundary seen by an
external observer is the same as that seen by an internal observer. A
more physical proof of this result is the following. We consider the
propagation of light emitted with frequency $\nu_0$ at the origin and
directed toward an observer outside the universe, at rest at
$\rho\,{=}\,\rho_{obs}\,{=}\,const$. According to the
eqs.~(\ref{eq:16f}) and (\ref{eq:ex.1}), on its whole way inside and
outside the universe the propagation of light is described by the
equation $\dot{\rho}(\tau)\,{=}\,c$ or $\rho\,{=}\,c\,(\tau{-}\tau_0)$
respectively. In consequence the time between the arrival of
successive wave crests at the place of the observer in ``outer space''
equals the time between their emission at the origin whence
$\nu_{obs}\,{=}\,\nu_0$. An observer comoving with the boundary of the
universe sees the origin receding from him at the velocity $v$ given
by the second of eqs.~(\ref{eq:x}). He therefore observes a redshift
of the light from the origin, and according to eq.~(\ref{eq:17-}) he
measures the frequency%
\vskip -\baselineskip
\begin{displaymath}
  \nu = \nu_0\,\left(\frac{1-v/c}{1+v/c}\right)^{1/2}\,.
\end{displaymath}
The light leaves him with the same frequency as it had on its
arrival. Since the boundary moves toward the external observer, the
latter will observe a blueshift of the light from the boundary and
measure the frequency
\begin{displaymath}
  \nu_{obs} = \nu\,\left(\frac{1+u/c}{1-u/c}\right)^{1/2}
    = \nu_0\,\left(\frac{1-v/c}{1+v/c}\right)^{1/2}\,
    \left(\frac{1+u/c}{1-u/c}\right)^{1/2}
\end{displaymath}
where $u$ is the velocity of the boundary in his (external) coordinate
frame. As we said above we must have $\nu_{obs}\,{=}\,\nu_0$, and with this
the last equation yields
\begin{equation}                                   \label{eq:extv}
  u = v = c\,\tanh\frac{\alpha\,R}{c}
\end{equation}
where at last eq.~(\ref{eq:x}) and $r\,{=}\,R$ was used.

\subsubsection{Numerical values of characteristic velocities}
1. In the comoving coordinates of the Robertson-Walker metric,
according to Hubble's law the present recessional velocity of the
boundary of the observable universe is $v \,{=}\, H_0\,d_0$ where
$d_0\,{=}\,a(t_0)\,r_{bo}\approx 3.5\,ct_0$ is its metric distance
from us and $t_0$ the age of the universe. With $H_0\,{=}\, 0.7\cdot
3.24\cdot 10^{-18}\,\mbox{s}^{-1}$ it becomes
$v_{rec}\,{=}\,3.45\,c$. A reasonable assumption about the outer
boundary $R$ of the universe is that at present it is two times as far
away from us as the present boundary of the observable universe,
i.e. $d\approx 7\,ct_0$. (This way inhomogeneities propagating from
``outer space'' into the universe cannot have spoiled the homogeneity
and isotropy inside the observable universe, see~\cite{Mukhanov}, page
231). The corresponding recessional velocity is%
\vskip -1.5\baselineskip
\begin{equation} 
  v_{rec} = 6.9\,c\,.
\end{equation}

2. In explosion coordinates the velocity at which the present boundary
of the observable universe moves away from us is given by the second
of the eqs.~(\ref{eq:x}). Following from $a(t) \,{=}\, \alpha\,t$ we
have
$\alpha\,r_{bo}/c\,{=}\,a(t_0)\,r_{bo}/(ct_0)\,{=}\,d_0/(ct_0)\,{=}\,3.5$,
and with this we obtain $ v_{bo} \,{=}\, 0.998\,c$.  At the outer
boundary of the universe we have $\alpha\,R/c\,{=}\,d/(ct_0)\,{=}\,7$,
and according to eq.~(\ref{eq:extv}) the velocity at which it moves is%
\vskip -\baselineskip
\begin{equation} 
  u = v = 0.999998\,c\,.
\end{equation}

\subsection{Universe with constant expansion rate}

We consider a second case in which a full analytic solution for
explosion coordinates can be obtained, namely
$a(t)\,{=}\,\alpha\,\mbox{e}^{Ht}$ with constant expansion
rate~$H$. In this case which describes an \textit{inflationary
  expansion}, in addition to the Doppler shift there is a
gravitational blueshift of the light from far away galaxies.

\subsubsection{Derivation of explosion coordinates}
In the present case, instead of using eqs.~(\ref{eq:8d}) and
(\ref{eq:8e}) it is easier to go back to eq.~(\ref{eq:6}), i.e.%
\vskip -\baselineskip
 \begin{displaymath}
   t_{\rho\rho} +H\,t_{\rho} = - H/c^2 \,.
 \end{displaymath}
 After solving the homogeneous equation, the solution of the
 inhomogeneous equation can be obtained by variation of constants and
 is given by
\begin{equation}                                     \label{eq:19}
  t(\rho,\tau) = \tau +\frac{1}{H}\,\ln\cos w \quad\quad\mbox{with}\quad\quad
  w = \frac{H\,\rho}{c}\,.
\end{equation}
An integration function was chosen such as to satisfy the second of the
conditions~(\ref{eq:9}). With this result and
$a(t)\,{=}\,\alpha\,\mbox{e}^{Ht}$ from eq.~(\ref{eq:8e}) we get
\begin{equation}                                           \label{eq:20} 
   r(\rho,\tau) = \frac{c\,\mbox{e}^{-H\tau}}{\alpha\,H}\,\tan w\,.
\end{equation}
Inversion of the functions $r(\rho,\tau)$ and $ t(\rho,\tau)$ given in
eqs.~(\ref{eq:19})--(\ref{eq:20}) yields
 \begin{equation}
   \label{eq:21}
   \rho = \frac{c}{H}\arcsin\left(\frac{\alpha\,H\,r}{c}\,\mbox{e}^{Ht}\right)\,,
   \quad\quad
   \tau = t - \frac{1}{2H}\,\ln\left(1-\frac{\alpha\,H^2\,r^2}{c^2}
     \,\mbox{e}^{2Ht}\right)\,.
 \end{equation}
 Inserting $a(t)\,{=}\,\alpha\,\mbox{e}^{Ht}$ and
 eqs.~(\ref{eq:19})--(\ref{eq:20}) in eq.~(\ref{eq:8b}) and the fourth
 of the eqs.~(\ref{eq:4}) we obtain%
\vskip -0.5\baselineskip
\begin{equation}                                      \label{eq:22}
  g_{00} = \cos^2 w\,,\quad\quad 
  g_\Omega = -\frac{c^2}{H^2}\,\sin^2 w\,.
\end{equation}
From the first equation it follows that eq.~(\ref{eq:8bb}) is
satisfied as it should according to the general theory.

\subsubsection{Flight velocity of the cosmic substrate  
  and associated redshift}
From the eqs.~(\ref{eq:8g}), (\ref{eq:19}) and
$a(t)\,{=}\,\alpha\,\mbox{e }^{Ht}$ we obtain the flight velocity
\begin{equation}                                         \label{eq:24}
  v = c\,\sin w\,.
\end{equation}

1. In the present case eqs.~(\ref{eq:8h}) and (\ref{eq:8hh}) become
\begin{displaymath}                                         
  \frac{\lambda_0}{\lambda_{em}} = \mbox{e}^{H(t_0-t_{em})}
  \quad\mbox{with}\quad 
  r_{em}=\frac{c}{\alpha}\int_{t_{em}}^{t_0} \mbox{e}^{-H\,t}dt
  =\frac{c}{\alpha\,H}\left(\mbox{e}^{-H\,t_{em}}-\mbox{e}^{-H\,t_0}\right)
\end{displaymath} 
and in combination yield
\begin{displaymath}                                     
  \frac{\lambda_0}{\lambda_{em}} 
  = \left(1-\frac{\alpha\,H\,r_{em}}{c}\,\mbox{e}^{H\,t_{em}}\right)^{-1}\,.
\end{displaymath} 
From this, with $r_{em}\to r$, $t_{em}\to t$ and
$\alpha\,H\,r/c\,{=}\,\mbox{e}^{-H\tau}\,\tan w$ according to
eq.~(\ref{eq:20}) and using eq.~(\ref{eq:19}), for
expansion coordinates we obtain the result
\begin{equation}                                          \label{eq:25}
  \frac{\lambda_0}{\lambda_{em}} 
  = \frac{1}{1-\sin w}\,.
\end{equation} 

2. In the system of explosion coordinates, from eqs.~(\ref{eq:17-}) 
and (\ref{eq:24}) we obtain
\begin{equation}                                      \label{eq:26}
  \frac{\lambda_\rho}{\lambda_{em}}=\frac{1+\sin w}{\cos w}\,,
\end{equation}
and from eq.~(\ref{eq:8ccd}) and the first of the eqs.~(\ref{eq:22})
we get
\begin{equation}                                      \label{eq:27}
  \frac{\lambda_0}{\lambda_\rho}=\sqrt{\frac{\cos^20}{\cos^2w}}=\frac{1}{\cos w}\,.
\end{equation}
Combing the results (\ref{eq:26}) and (\ref{eq:27}) we obtain the same
result (\ref{eq:25}) as in the system of expansion coordinates.

\subsubsection{Volume expansion rate}
According to eq.~(\ref{eq:22}) the volume of a spherical shell of
small thickness $\Delta\rho$ is given by
\begin{displaymath}
  \Delta V(\rho,\tau)= 4\pi\,\frac{c^2}{H^2}\,\Delta\rho\,\sin^2\!w\,.
\end{displaymath}
Because it is time independent, the expansion rate $E\,{=}\,d\Delta
V/(\Delta V\,\sqrt{g_{00}}d\tau)$ vanishes everywhere. This means that
in the present case, exponential inflationary evolution, the cosmic
substrate is moving relative to a completely invariable space.

\subsubsection{Acceleration of the cosmic substrate and comparison
  with ordinary explosions}

From the first of the eqs. (\ref{eq:21}) and from eq. (\ref{eq:24})
with
\begin{displaymath}
  \left.\frac{\partial \rho}{\partial
      t}\right|_r=c\,\tan\frac{H\,\rho}{c}
  = c\,\tan w
\end{displaymath}
for simplicity expressed in terms of the time $t$ we obtain
\begin{displaymath}
  \dot{v}(t) =  c\,\cos w \,\left.\frac{\partial w}{\partial t}\right|_r 
  = H\,\cos w \,\left.\frac{\partial \rho}{\partial t}\right|_r 
  = c\,H\,\sin w = H v\,.
\end{displaymath}
From this there follows an exponentially growing acceleration of the
cosmic substrate along its radial trajectories. Accordingly the
inflation preceding an almost big bang describes the actual explosion
whereas the subsequent dynamical processes are basically consequences
of it. (In the case of a big bang without inflation the dynamics of
explosion is compressed into a singular instant.) Like in ordinary
explosions the accelerated substance -- here dark energy, there e.g. a
chemical explosive like dynamite or gunpowder -- simultaneously is the
blasting agent driving the explosion, and for an external observer it
moves outward filling a previously empty space. What differs from
ordinary explosions is that there are no shattered fragments of a
casing, and there is no precursive shock wave. Instead there will be a
precursive front of the inflation field, a weak discontinuity, which
propagates at the speed of light.

\subsection{Hubble Law}

In expansion coordinates the distance from the origin of a galaxy or
an element of the cosmic substrate is $d\,{=}\,a(t)\,r$. From this
follows the Hubble law
\begin{equation}                                            \label{eq:Hub}
  \dot{d}(t) = H\,d(t) \quad\quad\mbox{with}\quad\;
  H = \frac{\dot{a}(t)}{a(t)}\,.
\end{equation}
In the explosion coordinates belonging to the case of constant
expansion velocity, $a(t)\,{=}\,\alpha\,t$, according to the second of the
eqs.~(\ref{eq:x}) we can write the flight velocity of galaxies etc. in
the forms%
\vskip -1.5\baselineskip
\begin{equation}                                            \label{eq:Hub2}
  v = H_{expl}\,\rho  \quad\quad\mbox{with}\quad\; 
  H_{expl} = \frac{1}{\tau}\,.
\end{equation}
Formally, this is identical with the Hubble law
\begin{displaymath}
   \dot{d}(t) = H\,d(t) \quad\quad\mbox{with}\quad\;
  H = \frac{1}{t}
\end{displaymath}
obtained in expansion coordinates. Physically the two laws are
different in that the underlying lengths as well as proper times are
measured differently.

Let us now consider the case of constant expansion rate $H$. In
expansion coordinates eq.~(\ref{eq:Hub}) with $H\,{=}\,const$
holds. In explosion coordinates according to eq.~(\ref{eq:24}) we have
\begin{equation}                                         \label{eq:Hex}
  v = H_{expl}\,\rho  \quad\quad\mbox{with}\quad\; 
  H_{expl} = \frac{c\,\sin(H\rho/c)}{\rho}
  = H\left(1-\frac{H^2\rho^2}{6c^2}+\dots\right) \,,
\end{equation}
where at last $ H_{expl}$ was expanded with respect to $H\rho/c$. In
contrast to the result for expansion coordinates, $H_{expl}$ is weakly
space dependent. This is not in contradiction to the result obtained
for expansion coordinates or from observations although a
space dependent Hubble parameter appears unusual. The reason is that
measurements concerning far away objects involve the application of a
theory related to the specific coordinate system in use, and this
theory must appropriately be adjusted in the transition to
explosion coordinates. (For example, the luminosity distance must be
redefined.)

\section{Applications and extensions}
\label{sec:applic}

1. In more general cases than the ones considered in the last section
$g_{00}$ will depend on $\rho$ and~$\tau$. Therefore, in addition to
the Doppler shift a time dependent gravitational redshift of light
will occur. While the Doppler shift can still be calculated from the
Doppler formula~(\ref{eq:17-}), no generally valid formula for the
shift effect of time dependent gravitation is known (at least to the
author of this paper). However, for all cases covered by
eq.~(\ref{eq:2}) the latter can be derived by making use of the
equivalence of the two different conceptions of galaxy recession.
Combining eqs.~(\ref{eq:8h}), (\ref{eq:8hh}) and (\ref{eq:17}), with
$t_{em}\to t$ as in eq.~(\ref{eq:17}) we obtain
\begin{equation}
   \frac{\lambda_0}{\lambda_\rho}
   =  \frac{\lambda_0}{\lambda_{em}} \frac{\lambda_{em}}{\lambda_\rho}
   = \frac{a(t_0)}{a(\tau)\,\left(1{+}\sqrt{1{-}a^2(t)/a^2(\tau)}\right)} 
   \quad\mbox{with}\quad
   r = r_{em}= c\int_{t_{em}}^{t_0}\frac{dt'}{a(t')}\,.
\end{equation}

2. Objects that are small in relation to cosmic distances like atoms,
the solar system or even galaxies do not participate in the spatial
expansion of the universe associated with usual theory. In
expansion coordinates this phenomenon is not easily comprehensible. A
first important proof of it is implicitly contained in one of
A.~Einstein's last papers~\cite{Einstein}. It is shown there that a
star in static equilibrium can smoothly be embedded in an expanding
universe which means that the radius of the star does not expand. From
this it is often derived that cosmic expansion is restricted to
distances of cosmic scale. (According to~\cite{Wheeler} ``only
distances between clusters of galaxies and greater distances are
subject to the expansion''.)  What cannot be understood on this basis
is, why, on the other hand, light rays of much smaller spatial
extension are still subject to expansion. (This applies in particular
to the incoherent light emitted by far away galaxies etc.. However, in
the usual derivation of the cosmic redshift only monochromatic wave
trains of infinite extension are considered.)  This different behavior
is especially difficult to understand in
expansion coordinates~\cite{Rebhan}, but it is very easily understood
in explosion coordinates: Completely independent of the radial
extension of a radially directed light ray the frequency of it is
redshifted due to the Doppler effect and the gravitational field.

3. The transformation from expansion  to explosion coordinates can
only in special cases be analytically expressed and will in general
involve numerical calculations. A formulation of the general
relativistic equations for the dynamic evolution of the universe in
explosion coordinates may offer an easier approach to solutions, at
least in special cases.

4. The introduction of explosion coordinates in a universe with
negative spatial curvature ($k\,{=}\,{-}1$) is certainly feasible in a similar
way as in the case $k\,{=}\,0$ considered in this paper. In the case of
positive spatial curvature the situation is different. Nevertheless it
may be possible to successfully impose the condition $g_{\rho\rho}\,{=}\,0$
also in this case, although this would appear somewhat artificial.

\section{Conclusion}
\label{sec:conclusion}

It was shown in this paper that explaining the recession of galaxies
(or other manifestations of matter at earlier stages) by an expansion
of space or as an explosion  and after-explosionlike motion relative
to space is equivalent. Both interpretations are relative in that
their validity is restricted to specific coordinate systems. The
transition between them can be performed by simple transformations.

Certainly, the usual approach using expansion coordinates has the
major merits.  Most important is its simplicity provided by the fact
that many physical properties are related to one simple time dependent
parameter, the scale factor~$a(t)$.  Furthermore, the coordinate time
is equal to the proper time and is thus valid for the whole
universe. For people unfamiliar with the field the concept of an
expanding space may occasionally provide difficulties, but this is
by far surpassed by the afore-mentioned advantages.

Our approach by explosion coordinates aimed at an alternative
interpretation and consequently is based on a reformulation of known
results. Nevertheless it has its merits as well. For one, it better
fits the view of an external observer. Also, the interpretation of the
big bang or inflation as giant explosions, and the restriction of
recessional velocities to values below the velocity of light are more
intuitive. Furthermore there exist problems which are easier to handle
in explosion coordinates. Finally, the close affinity to Newtonian
cosmology may provide advantages in some cases.

The two interpretations should by no means be confounded. As well as
the Doppler effect must not be used for explaining the redshift in
expansion coordinates, their corresponding distances and recessional
velocities must not be employed when Doppler effect and gravitational
blueshift are evaluated in explosion coordinates.

\end{document}